\documentstyle[PASJadd]{PASJ95}
\begin{document}
\setcounter{page}{001}

\title{Detections of SiO and H$_2$O Masers 
in the Bipolar Nebula\\ IRAS 19312+1950}

\author{Jun-ichi {\sc Nakashima}\\
{\it Department of Astronomical Science, The Graduate University for 
Advanced Studies,}\\
{\it Nobeyama Radio Observatory, Minamimaki, Minamisaku, Nagano 384-1305} 
\\
{\it E-mail(JN): junichi@nro.nao.ac.jp}\\[6pt]
 and\\  
 Shuji {\sc Deguchi} \\
{\it Nobeyama Radio Observatory, National Astronomical Observatory}\\
{\it Minamimaki, Minamisaku, Nagano 384-1305}\\
}

\abst{
We report on the detection of SiO and H$_2$O masers toward a newly found bipolar nebula, IRAS 19312+1950. This object exhibits extreme red IRAS color [$\log (F_{25}/F_{12})=0.5$ and $\log (F_{60}/F_{25})=0.7$] and a nebulosity having a size of about 30$''$ extended to the South--West in the 2MASS near-infrared image. Toward this object, we have detected emission from the H$_{2}$O 6(1,6)--5(2,3) transition, the SiO $J=1-0, v=1$ and 2, and $J=2-1, v=1$ transitions, and the SO 2(2)--1(1) and H$^{13}$CN  $J=1-0$ transitions. The thermal lines of SO and H$^{13}$CN are shifted by about 12 km s$^{-1}$ in radial velocity with respect to the maser lines, indicating that thermal emission comes from the background molecular cloud. However, the SiO $J=2-1, v=2$ spectrum shows another component of SiO emission separated by 26 km s$^{-1}$ from the main component, that might be formed in a rotating or expanding shell.
}

\kword{ISM: molecules --- masers --- planetary nebulae: individual(IRAS 19312+1950 --- stars: AGB and post--AGB --- stars:mass--loss}

\maketitle
\thispagestyle{headings}

\section
{Introduction}

SiO, H$_2$O, and OH masers have been known to be associated with late-type stars. These are stars on the Asymptotic Giant Branch (AGB) with a large mass loss rate. Occasionally, an OH 1612 MHz maser has been found in the envelopes of further evolved objects, e.g., proto-planetary and planetary nebulae (Nyman et al. 1998). The central stars of these nebulae are in the post-AGB phase or beyond (white dwarf), in which the AGB mass loss has already ceased. 

These evolved objects are very often recognized based on their peculiar shape on the optical and near-infrared images (Kwok 2000), or from their spectral energy distribution(Kwok 1990). The point-symmetric morphology of proto-planetary and planetary nebulae has been well studied recently with the Hubble Space Telescope (Sahai 1999; Ueta et al. 2000). However, it is not clear whether the aspherical shape of these objects is developed throughout the AGB phase, or only in the last transient phase of mass loss to the proto-planetary nebula. 

SiO masers are emitted in an envelope close to the central star, and OH masers much further away (Lewis 1989). When the mass loss ceases, the H$_2$O and SiO masers disappear within a few years, whereas the OH masers remain active also during the proto-planetary and early planetary nebula stages (Lewis 1989). Therefore, SiO and H$_2$O masers can be useful probes for the mass loss of transient objects.

During a survey of galactic disk objects with very red IRAS colors for SiO masers at Nobeyama, we found SiO masers in IRAS 19312+1950 with $F_{12}=22.5$ Jy and the colors $C_{12}\equiv \log (F_{25}/F_{12})=0.50$ and $C_{23}\equiv \log (F_{60}/F_{25})=0.70$, one of the reddest objects with SiO masers in this survey. Here, $F_{12}$, $F_{25}$, and $F_{60}$ are the IRAS flux densities at 12, 25, and 60 $\mu$m, respectively. A bipolar-type nebulosity is seen in the 2MASS near-infrared, $J$,$H$,$K$--band image of this object. Up to now, one positive detections and three questionable detections of SiO masers from proto-planetary nebula candidates have been reported (Nyman et al. 1998). This object is the second detection of SiO masers in this type of object. The observed spectra of maser and thermal lines toward IRAS 19312+1950 show slightly odd characteristics, i.e., two peaks in the SiO maser spectrum, and a shift in the radial velocity of the H$^{13}$CN and SO lines from the maser lines. In this letter, we report on molecular line observations of this newly found interesting object.
\par
\vspace{1pc}\par

\section{Observations}
The observations were made in 2000 May with the 45-m radio telescope at Nobeyama. The detected lines and the observed line parameters are summarized in table 1. The galactic coordinates of IRAS 19312+1950 are ($l$, $b$) = (55.4$^{\circ}$, 0.20$^{\circ}$). The IRAS position at the epoch of 1950.0 was used for the observations. The half-power beam width (HPBW) of the telescope was 72$''$ at 23 GHz, 38$''$ at 43 GHz and 18$''$ at 86 GHz, respectively. The pointing was checked every 2 or 3 hours with the nearby strong SiO maser sources, V111 Oph, or, $\chi$ Cyg. A typical position error for an IRAS source is less than 10$''$ (Jiang et al. 1996), accurate enough if compared with the telescope HPBW of about 20$''$ at 43 GHz. The typical pointing accuracy of the telescope was about 5$''$, or under windless condition, a pointing accuracy better than 2$''$ was achieved.  We used a cooled HEMT amplifier at 22 GHz, and SIS mixer receivers at 43 and 86 GHz. The system temperature (including atmospheric noise) was 190--400 K at 22 GHz, 160--270 K at 43 GHz, and 250--300 K at 86 GHz, depending on the weather conditions. The aperture and main-beam efficiencies were 0.64 and 0.82 at 23 GHz, 0.58 and 0.73 at 43 GHz, and 0.45 and 0.53 at 86 GHz, respectively. Acousto--optical spectrometer arrays,  AOS-H and AOS-W,  with 40 and 250 MHz bandwidths with 2048 frequency channels each were used as backends. A high-resolution spectrometer (AOS-H) was used for the SiO ($J=1-0, v=1$ and 2)  and H$_{2}$O maser lines, and a spectrometer with a wide band coverage (AOS-W) was used for both SiO ($J=2-1, v=1$ and $J=1-0, v=1$ and 2) lines. The SO 2(2)--1(1) and H$^{13}$CN $J=1-0$ lines at 86.094 and 86.340 GHz were also observed with the same spectrometer (AOS-W) observing the SiO $J=2-1, v=1$ line. The effective spectral resolution was 1.74 km s$^{-1}$ at 43 GHz and 0.87 km s$^{-1}$ at 86 GHz in the case of AOS-W, and 0.28 km s$^{-1}$ at 43 GHz, and 0.55 km s$^{-1}$ at 22 GHz in the case of AOS-H. All of the observations were made in the position--switching mode using a 10$'$ off position in azimuth. 

The observational results are summarized in table 1 and the spectra obtained are shown in figure 1. Emission from the SiO $J=2-1, v=1$, and $J=1-0, v=1$ and 2, and H$_{2}$O lines were found near $V_{\rm lsr}=20$ km s$^{-1}$. However, the SO and H$^{13}$CN  lines ($V_{\rm lsr} \simeq 32$ km s$^{-1}$) are significantly shifted in radial velocity from the maser lines. Looking carefully at the spectrum of the SiO $J=1-0, v=2$ transition (the second spectrum from the bottom in figure 1), we recognized the presence of another component at 52.8 km s$^{-1}$. The corresponding component in the SiO $J=1-0, v=1$ spectrum was missing. In the spectrum of the SiO $J=1-0, v=1$ transition, a faint secondary component can be seen around $V_{\rm lsr}=18.0$ km/s, at the left of main peak at 25.5 km s$^{-1}$. These differences in radial velocities in the different transitions need careful interpretation, and are discussed in the next section.
\par
\vspace{1pc}\par

\section{Discussion}

The IRAS color of 19312+1950,  $(C_{12}, C_{23})=(0.50, 0.78)$, suggests a uniqueness of this object.  In a two-color diagram (van der Veen, Habing 1988), this object falls into region VIII,  which is occupied by a number of very cool stars. About 30\% of objects with the IRAS LRS spectra in region VIII indicate planetary nebulae. The LRS spectra of this object (Volk et al. 1991) exhibits weak silicate absorption on a red continuum. On the other hand, this object mimics the color of young stars found in star--forming regions (MacLeod et al. 1998a). Based on the IRAS colors, it is hard to classify IRAS 19312+1950 as a young planetary nebula or a young star in a star--forming region. It is not surprising that a negative result of the 6.7 GHz methanol maser search was reported toward this object (MacLeod et al. 1998b). 

Figure 2 shows the 2MASS near-infrared image (Skrutskie et al. 2000) \footnote{Atlas Image obtained as part of the Two Micron All Sky Survey (2MASS),  a joint project of the University of Massachusetts and the Infrared Processing  and Analysis Center/California Institute of Technology, funded by the National Aeronautics  and Space Administration and the National Science Foundation, U.S.A.} of the region around IRAS 19312+1950. We can see a nebulosity with two horns extended towards to south west. These horns can clearly be seen on this image, though the IRAS source is point-like in the $J$ band. From this image and the detections of SiO masers in this source, we conclude that this is a post-AGB star surrounded by a bipolar-type nebulosity (like the Rotten Egg Nebula, OH 231.8+4.2; Morris et al. 1987). There is no nearby infrared star which is bright in the 2MASS $K$-band image. We conclude that the extended nebulosity at the center of figure 2 is a part of this object.

A more observational investigation is necessary to clarify the nature of this object. For instance, the radial velocities of the maser and thermal lines toward this object are puzzling. From the velocities of the maser lines (SiO and H$_{2}$O), we infer a stellar velocity of approximately $V_{\rm lsr}=20$ km s$^{-1}$ [it is well known that the SiO radial velocity coincides with the stellar velocity (Jewell et al. 1991), while H$_2$O spectra often exhibit single or double peaks  centered at the stellar velocity (Engels, Lewis 1996)]. However, the $J=1-0, v=2$ spectrum exhibits another component at  $V_{\rm lsr}=52$ km s$^{-1}$. It is possible that this interloper component comes from a nearby source which is located by chance in the same telescope beam of about 40$''$ toward this object; such examples are already known for two cases among the densely populated disk IRAS sources (Deguchi et al. 1999). On the other hand, it is possible to interpret this component as part of a doubly peaked profile which is formed in a rotating disk or expanding shell surrounding this object. In this case, the stellar velocity would be $V_{\rm lsr} \simeq 39$ km s$^{-1}$. If this is correct, it is also possible (but less plausible) to interpret this source as a young stellar object. In past studies, the detection of SiO masers in several star-forming regions has been reported (e.g., Hasegawa et al. 1986; Snyder, Buhl 1974). A doubly peaked profile with a separation of more than 20 km s$^{-1}$ was found in Orion IRc2, and has been explained by a rotating and expanding disk model (Plambeck et al. 1990; Barvainis 1984). On the other hand, SiO masers from proto-planetary nebula show a somewhat broad line profile in past studies; it resembles to a typical profile of SiO masers from a circumstellar envelope of the Mira-type variables (e.g., Nyman et al. 1998). Up to now, no clear doubly peaked profile of SiO masers in the bipolar nebulae has been reported. 

The radial velocities of H$^{13}$CN and SO are shifted by about 12 km s$^{-1}$ from the maser lines. Because of this large separation in radial velocities and the slightly asymmetric profiles of these lines, we consider that these thermal lines come from a background molecular cloud. According to the Columbia CO survey in the galactic disk (Cohen et al. 1986), CO emission at $(l, b)= (55.0^{\circ}, 0.25^{\circ})$ is peaked at $V_{\rm lsr}\simeq$ 30 km s$^{-1}$, which coincides with the radial velocities of H$^{13}$CN and SO found in this paper. However,  we cannot completely neglect the possibility that the two broad lines of H$^{13}$CN and SO come from the outflowing envelope of this object, unless the component ($V_{\rm lsr}=$52 km s$^{-1}$) of SiO is proved to come from a totally different source by chance, and until the CO emission in the background molecular cloud is mapped by telescopes with better spatial resolution.

It is known that very few sources are detected in SiO masers beyond $C_{12}=0.5$ (Nyman et al. 1998); only two stars, OH 231.8+4.5 and OH 15.7+0.8, are such extreme sources with $C_{12}>0.5$ accompanying SiO. However, the color $C_{23}$ of these sources are below 0.4. One of the sources with the colors compatible to 19312+1950, is IRAS 18498$-$0017 (OH 32.8$-$0.3; $C_{12}=0.46$ and $C_{25}=0.74$); H$_2$O, SiO, and OH masers have been detected in this source (Cesaroni et al. 1988; Engels, Heske 1989; te Lintel Hekkert et al. 1989). Near-infrared images of this source are not available at present in the 2MASS archive. It is quite interesting to see if this source also exhibits a bipolar-type nebulosity or not. The color, $C_{23}\sim 0.7$, of these two objects seems to be meaningfully higher than those of OH 231.8+4.5 and OH 15.7+0.8. However, because the positions of 18498$-$0017 and 19312+1950 are very near to the galactic plane ($b=-0.04^{\circ}$ and 0.2$^{\circ}$), emission from the interstellar dust grains possibly contaminates the 60 $\mu$m flux densities of these objects, as found by the Nobeyama SiO maser survey of the galactic disk IRAS sources (Izumiura et al. 1999). The long-term near-infrared  monitoring of 18498$-$0017  (Engels et al. 1983) found a variability  with a time-scale of more than 1700 d, suggesting that 18498$-$0017 is not a normal Mira variable.  It may be at a transient stage to the proto-planetary nebula.
\par
\vspace{1pc}\par

\section
{Summary}
In this letter, we have reported the detection of SiO and H$_2$O masers from a newly found bipolar nebula, IRAS 19312+1950. We also observed thermal lines of SO and H$^{13}$CN toward this source, though they are shifted in radial velocity and are probably attributed to the molecular cloud in the same direction.

The authors thank Dr. M. Parthasarathy for stimulating discussions. This research was partly supported by a Scientific Research Grant (C) 20197825 from the Japan Society of Promotion of Sciences.
\par
\vspace{1pc}\par
\vspace{1pc}\par

\newpage

\begin{fv}{1}{25pc}%
{
Spectra of the molecular lines detected in IRAS 19312+1950. The antenna temperatures of the bottom three spectra are scaled by factors of 2, 2, and 0.15. The vertical broken line shows a guide for the eye to mark the stellar radial velocity which, is adopted to be at about $V_{\rm lsr}=$20 km s$^{-1}$. The spectrum in the SiO $J=1-0 v=1$ transition was contaminated by strong spurious signals in the receiver system at $V_{\rm lsr}=-10$ --- $-5$ km s$^{-1}$, so that the bad channels were removed.
}
\end{fv}

\begin{fv}{2}{20pc}%
{
Composite near-infrared image from 2MASS data archive, composed of $J$, $H$ and $K$-band images. The image size is 1$'\times 1'$. The center of the image is the IRAS position of 19312+1950, and north is up, west is right.
}
\end{fv}

\newpage

\begin{table*}
\renewcommand{\baselinestretch}{1.0}
\small
\begin{center}
Table~1.\hspace{4pt}Observational Results.\\
\end{center}
\vspace{6pt}
\begin{tabular*}{\textwidth}{@{\hspace{\tabcolsep}
\extracolsep{\fill}}p{4pc}cccccccc}
\hline\hline\\[-6pt]
Molecule & Transition & Frequency & T$_{\rm peak}$ & V$_{\rm lsr}$ 
& Flux & r.m.s. & obs. date \\
[4pt]\hline\\[-6pt]
 &  & (GHz) & (K) & (km s$^{-1}$) & (K km s$^{-1}$) & (K) & (yyyy.mm.dd) \\
H$_{2}$O & 6(1,6)-5(2,3) &22.2351 & 2.339 &17.4 &7.455 & 0.127 & 
2000.05.30 \\
 &  &  &  &  &  &  &  \\
SiO & $1-0$ $v$=2 & 42.8206 & 0.118 & 26.8 & 0.357 & 0.022 & 2000.05.30 \\
   &   &  & 0.107 & 52.8 & 0.345 & 0.022 & 2000.05.30 \\
 &  &  &  &  &  &  &  \\
SiO & $1-0$ $v$=1 & 43.1221 & 0.113 & 25.5 & 0.607 & 0.027 & 2000.05.30 \\
 &  &  &  &  &  &  &  \\
SO & $2(2)-1(1)$ & 86.0936 & 0.149 & 32.0 & 1.657 & 0.015 & 2000.05.25 \\
 &  &  &  &  &  &  &  \\
SiO & $2-1 v=1$ & 86.2434&0.285 & 17.4 & 1.962 & 0.015 & 2000.05.25 \\
 &  &  &  &  &  &  &  \\
H$^{13}$CN & $1-0$ ($F=2-1$) & 86.3402 & 0.165 & 32.3 & 1.110 & 0.019 & 
2000.05.25 \\

\\[4pt]
\hline
\end{tabular*}
\vspace{6pt}
\noindent
\end{table*}

\newpage
\newpage
\section*{References}
\small

\re
Barvainis R.\ 1984, ApJ 279, 358

\re
Cesaroni R., Palagi F., Felli M., Catarzi M., Comoretto G., Di Franco S., 
Giovanardi C., 
Palla F.\ 1988, A\&AS 76, 445

\re
Cohen R.S., Dame T.M., Thaddeus P.\ 1986, ApJS 60, 695

\re
Deguchi S., Fujii T. Izumiura H., Matsumoto S., Nakada Y., Wood P. R.,    
Yamamura I.\ 1999,  PASJ 51, 355

\re
Engels D., Heske A.\ 1989, A\&AS 81, 323

\re
Engels D., Kreysa E., Schultz G.V., Sherwood W.A.\ 1983, A\&A 124, 123

\re
Engels D., Lewis B.M.,\ 1996, A\&AS 116, 117

\re
Hasegawa T., Morita K.-I., Okumura S., Kaifu N., Suzuki H., Ohishi M., 
Hayashi M., Ukita N.\ 1986, in Masers, Molecules and Mass Outflows in Star 
Forming Regions, ed Haschick A.D. (Haystack Observatory, MA) p275

\re
Izumiura H., Deguchi S., Fujii T., Kameya O., Matsumoto S. Nakada Y., 
Ootsubo T., Ukita N., 1999,  ApJS, 125, 257

\re
Jewell P.R., Snyder L.E., Walmsley C.M., Wilson T.L., Gensheimer P.D.\ 
1991, A\&A 242, 211

\re
Jiang B.W., Deguchi S., Nakada Y.\ 1996, AJ 111, 231

\re
Kwok S.,\ 1990, A.S.P. Conf. Ser. 9, 438

\re
Kwok S., 2000, The Origin and Evolution of Planetary Nebulae (Cambridge 
Univ. Press., Cambridge), p.89

\re
Lewis B.M.\ 1989, ApJ 338, 234

\re
MacLeod G.C., Scalise E.Jr., Saedt S., Galt J.A., Gaylard M.J.\ 1998a, AJ 
116, 1897

\re
MacLeod G.C., van der Walt D.J., North A., Gaylard M.J., 
Galt J.A., Moriarty-Schieven G.H.\ 1998b, AJ 116, 2936

\re
Morris M., Guilloteau S., Lucas R., Omont A.\ 1987, ApJ 321, 888

\re
Nyman L.-\AA., Hall P.J., Olofsson H.\ 1998, A\&AS 127, 185

\re
Plambeck R.L., Wright M.C.H., Carlstrom J.E.\ 1990, ApJ 348, L65

\re
Sahai R.\ 1999, ApJ 524, L125

\re
Skrutskie M.F., Stiening R., Cutri R., Beichman C., Capps R., Carpenter J.,
Chester T., Elias J. et al.\ 2000 (The 2MASS Team)\\ 
http://www.ipac.caltech.edu/2mass/index.html

\re
Snyder L.E., Buhl D.\ 1974, ApJ 189, L31

\re
te Lintel Hekkert P., Versteege-Hensel H.A., Habing H.J., Wiertz M.\ 
1989, A\&AS 78, 399

\re
Ueta T., Meixner M., Bobrowsky M.\ 2000, ApJ 528, 861

\re
van der Veen W.E.C.J., Habing H.J.\ 1988, A\&A 194, 125

\re
Volk K., Kwok S., Stencel R.E., Brugel E.\ 1991, ApJS 77, 607

\label{last}

\vspace{1pc}\par
\vspace{1pc}\par
\underline{\it Note in added proof}: We took the $JHK$ images and spectra of this object with the Siding Spring 2.3-m telescope in July 2000 and found that emission from the horn features is continuum, and that the H$_2$ (2.122 $\mu$m) emission comes from the position a few arc second north of the central star.

\end{document}